\documentclass[a4paper,preprintnumbers,floatfix,pra,twocolumn,showpacs,notitlepage,longbibliography,nofootinbib]{revtex4-1}

\usepackage{amsmath, amssymb, amsthm, thm-restate, bm, physics, dsfont, nicefrac}
\usepackage{xcolor, graphicx, subfigure, float, appendix, booktabs, microtype}
\usepackage[colorlinks,citecolor=pgreen,linkcolor=ppink,urlcolor=pgreen]{hyperref}

\newcommand{\id}{\mathds{1}}
\renewcommand{\tr}[1]{\text{tr}\left( #1 \right)}

\newcommand{\overbar}[1]{\mkern 1.5mu\overline{\mkern-1.5mu#1\mkern-1.5mu}\mkern 1.5mu}
\newcommand{\mean}[1]{\langle #1 \rangle}

\newcommand{\lov}{\vartheta(\overbar{G})}
\newcommand{\eps}{\varepsilon}

\definecolor{pgreen}{RGB}{107, 161 ,168}
\definecolor{ppink}{RGB}{214, 125 ,137}

\begin{document}
\title{Uncertainty relations from graph theory}

\author{Carlos de Gois}\thanks{These authors contributed equally to this paper.}
\author{Kiara Hansenne}\thanks{These authors contributed equally to this paper.}
\author{Otfried Gühne}
\affiliation{Naturwissenschaftlich-Technische Fakultät, Universität Siegen, Walter-Flex-Stra\ss e 3, 57068 Siegen, Germany}

\begin{abstract}
    Quantum measurements are inherently probabilistic and quantum 
    theory often forbids to precisely predict the outcomes of 
    simultaneous measurements. This phenomenon is captured 
    and quantified through uncertainty relations. Although studied 
    since the inception of quantum theory, the problem of determining 
    the possible expectation values of a collection of quantum measurements 
    remains, in general, unsolved. By constructing a close connection 
    between observables and graph theory, we derive uncertainty relations 
    valid for any set of dichotomic observables. For arbitrary observables, we obtain an upper bound on the sum of the square of their expectation values. We furthermore show how this bound behaves when the observables are imprecisely calibrated.
    As applications, our results can be straightforwardly 
    used to formulate entropic uncertainty relations, separability criteria 
    and entanglement witnesses.
\end{abstract}

\maketitle

\section{Introduction}
    Physical systems are characterised through the results of measurements,
    and in quantum theory measurements are described by observables. 
    Mathematically, these are linear operators with real eigenvalues, each 
    of which represents a possible result of a measurement. Born's rule allows 
    to compute for each of these outcomes the probability of happening, 
    depending solely on the state and the observable in question. The 
    mathematical structure of quantum theory, however, enforces constraints 
    on the probabilities, for instance if a pair of non-commuting observables 
    is considered. These constraints frequently manifest themselves as uncertainty 
    relations \cite{Heisenberg_1927,Kennard_1927,robertson29,robertson34,wehner2010entropic,coles2017entropic,busch2014colloquium,schwonnek2018phd}. More generally, significant research effort
    has been devoted to the problem of determining the set of possible 
    expectation values for sets of observables \cite{toth2005entanglement,szymanski2018classification,schulte2008significance,gawron2010restricted,wehner2008higher}, which is of high relevance for various applications, such as quantum cryptography, entanglement characterisation 
    and quantum metrology.

    The main problem can be explained with a simple, yet fundamental example. Consider
    a single spin-$1/2$ particle (or qubit) and let $X,Y,Z$ denote the Pauli matrices \cite{pauli1927matrices}. 
    Then, it is well known that for any quantum state the relation
    \begin{equation}
    \ev{X}^2 + \ev{Y}^2 + \ev{Z}^2 \leq 1
    \label{eq-fundamental}
    \end{equation}
    holds \cite{feynman1957geometrical},
    but it also gives rise to the uncertainty relation 
    $\Delta^2(X) + \Delta^2(Y) + \Delta^2(Z) \geq 2$,
    where $\Delta^2(A) = \mean{A^2}-\mean{A}^2$ denotes the variance. This uncertainty
    relation is not only of fundamental interest, but it also found application in 
    entanglement characterisation \cite{hofmann2003violation} and quantum cryptography \cite{portmann2022security}. But how can this relation be generalised?
    
    To do that, it is natural to consider the expression 
    $\mathbb{E} = \sum_{i} \mean{A_i}_\varrho^2$, where each $A_i$ is 
    an observable with spectrum $\pm 1$, and $\varrho$ a quantum state.
    As a simple generalisation of Eq.~(\ref{eq-fundamental}) it is known 
    that, for any set of pairwise anticommuting observables, $\mathbb{E} \leq 1$ \cite{toth2005entanglement,wehner2008higher,wehner2010entropic}. This 
    inequality led to the derivation of entropic uncertainty relations for 
    multiple observables \cite{wehner2008higher,niekamp2012entropic}, Bell 
    monogamy inequalities \cite{kurzynski2011correlation} and quantum network 
    state compatibility criteria \cite{hansenne2022symmetries}, among others. 
    Moreover, it leads to an uncertainty relation for multiple observables, 
    $\sum_{i=1}^n \Delta^2(A_i) \geq n - 1$. Similar preparation uncertainty relations%
    \footnote{ We remark that at least two concepts of \emph{measurement uncertainty relations} also exist in the literature. When dealing with preparation uncertainty relations, one is ultimately interested in the spread of the probability distributions obtained from measuring the given observables on independent copies of some state $\varrho$. Conversely, measurement uncertainty relations take into account the disturbance introduced by a measurement onto another when they are performed simultaneously or sequentially. See \cite{ozawa2003universally} and \cite{busch2014heisenberg} for two competing approaches to measurement uncertainty relations.}
     have been previously studied \cite{maccone2014stronger,konrad2019geometric,abbott2016tight,schwonnek2017state,giorda2019state}, but the results were mainly 
    applicable to two or three observables and not to larger sets.

    \begin{figure}[t]
        \centering
        \includegraphics[width=.85\linewidth]{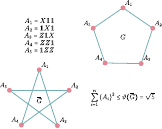}
        \caption{For any set $\{A_i\}_{i=1}^n$ of observables, we draw a graph $G$ with one vertex corresponding to each observable and an edge $\{i, j\} \in G$ if and only if the anticommutator $\{A_i, A_j\} \neq 0$. Its complement, $\overbar{G}$, is called anticommutativity graph, and the Lovász number $\lov$ provides an upper bound to $\mathbb{E}$.}
        \label{fig:summary}
    \end{figure}
    
    Generalisations of $\mathbb{E}$ to other sets of observables are thus scarce. One trivial fact is that $\mathbb{E} \leq n$, where $n$ is the number observables employed.  A more insightful fact is that, when $\{ A_i \}$ 
    contains $d^2$ operators that span the space of Hermitian operators acting 
    on a $d$-dimensional Hilbert space, with $\tr{A_iA_j}= d \delta_{ij}$, then $\mathbb{E} \leq d$. 
    This follows from the positivity of the density operator and can be interpreted 
    as an upper bound on the length of its Bloch vector \cite{kimura2003bloch,byrd2003characterization}. Further than combinations of these facts (for 
    instance, one may divide a set of observables into subsets, such that in
    each subset the observables anticommute), not much is known.

    In this paper we derive nontrivial bounds on $\mathbb{E}$ for {\it any} 
    set of observables by connecting this problem to graph theory. 
    More specifically, to any set of observables we associate a graph encoding
    the anticommutativity relations (see also Fig.\@ \ref{fig:summary}) and show 
    that  graph-theoretic quantities, such as the Lovász number or the order of a 
    maximum clique have a physical interpretation. For plenty of cases --- for 
    instance, for all perfect graphs --- the obtained bounds are tight. Then, we extend the bounds to the case of imprecise observables and show how they grow with respect to an imprecision parameter $\eps$. 
    Finally, we discuss how our method leads to a simple procedure for generating entanglement witnesses \cite{guhne2009entanglement} robust to imprecision, to generalisations of entropic uncertainty relations \cite{niekamp2012entropic,wehner2008higher,kaniewski2014entropic} and to applications 
    in nonlinear entanglement detection via local uncertainty relations \cite{hofmann2003violation,guhne2004characterizing}.

\section{Main connection}
    Without loss of generality, we will consider observables that square to the 
    identity, as any dichotomic observable can be rescaled to have a $\pm 1$-spectrum.
    To find an upper bound for $\mathbb{E}$, we start by defining $E$ as a linear combination of the given observables depending on a state $\varrho$, that is, 
    $E_\varrho = \sum_{i=1}^n a_i A_i$, with $a_i = \mean{A_i}_\varrho \in \mathbb{R}$. 
    The variance of this observable on the same state $\varrho$ is non-negative, 
    hence $\mean{E_\varrho^2} \geq \mean{E_\varrho}^2$ (here and later on, we will 
    omit the dependency of expectation values on $\varrho$). While the right-hand-side 
    is equal to $[ \sum_i \mean{A_i}^2]^2 = \norm{\Vec{a}}^2 \sum_i \mean{A_i}^2$, the l.h.s.\@ can be rewritten as
    \begin{equation}
            \sum_{ij} a_i a_j \mean{A_iA_j} = \sum_{ij: \{A_i,A_j\} \neq 0 } a_i a_j \mathfrak{R}\left(\mean{A_iA_j}\right) ,
    \end{equation}
    where we used that $\mean{A_jA_i} = \mean{A_iA_j}^*$. This sum is upper bounded by $\norm{\Vec{a}}^2 \Lambda(\mathcal{A})$, where $\Lambda(\mathcal{A})$ is the largest eigenvalue of the matrix $\mathcal{A}$ whose elements $\mathcal{A}_{ij}$ are $\mathfrak{R}(\mean{A_iA_j})$.

    Let us further specify $\mathcal{A}$. When $\{ A_i, A_j\} = 0$ (i.e., a pair of observables anticommute), then $\mathcal{A}_{ij} = 0$. As the observables are required to have $\pm 1$-spectra, they square to the identity, thus the diagonal elements of $\mathcal{A}$ are equal to one. Furthermore, the matrix $\mathcal{A}$ is real and positive semidefinite. So, for any state $\varrho$,
    \begin{subequations} 
    \label{eq:max-eigenvalue-bound}
        \begin{align}
        \sum\nolimits_{i=1}^n \mean{A_i}^2 \leq &\max\nolimits_{\mathcal{A}} \Lambda(\mathcal{A}) \\
                           &\text{ subject to:} \quad \mathcal{A} \in \mathbb{R}^{n \times n} \text{ and } \mathcal{A} \succeq 0, \\
                           &\quad \mathcal{A}_{ii} = 1, \,\forall i \in \{1, \dots, n\}, \label{eq:diagonals-mathcal-a-matrix}\\
                           &\quad \mathcal{A}_{ij} = 0, \,\forall \{i, j\}: \{A_i,A_j\}=0 .
        \end{align}
    \end{subequations}

    To establish the connection to graph theory, we define the \emph{anticommutativity graph} $\overbar{G}$ as the one with $n$ vertices 
    --- each corresponding to an observable --- connected by an edge if and 
    only if $\{ A_i, A_j \} = 0$, see also Fig.~\ref{fig:summary}. For general
    graphs, there are several quantities of interest. The {\it clique number} $\omega(G)$ 
    is the size of the largest subgraph of $G$ that is fully connected and the 
    {\it chromatic number} $\chi(G)$ is the smallest number of colours needed to colour 
    the vertices of $G$ such that no connected vertices share the same colour. 
    The {\it Lovász number}, also called $\vartheta$-function, was originally introduced 
    as an upper bound to the Shannon capacity of a graph \cite{lovasz1979shannon},
    and  can be defined in different equivalent ways (see also below). Remarkably, it 
    can also be used to estimate the previously mentioned quantities: The sandwich 
    theorem \cite{lovasz1979shannon} states that 
    $\omega(G) \leq \lov \leq \chi(G)$. Here, $G$ and $\overbar{G}$ are 
    complementary graphs, i.e. vertices which are connected in $G$ are disconnected
    in $\overbar{G}$ and vice versa. The sandwich theorem is of particular interest 
    since, whereas determining 
    $\omega(G)$ and $\chi(G)$ is an \textsc{np}-complete problem \cite{karp1972reducibility},
    $\lov$ can be computed through a semidefinite program \cite{lovasz1979shannon}. A graph
    $G$ for which these three quantities are equal for all its induced subgraphs is called a 
    perfect graph.
   
    Among the many, equivalent definitions of the Lovász number $\vartheta(G)$, 
    one is particularly interesting for our problem. As shown in Section 33 of \cite{knuth1994sandwich} (up to a typo in Eq.\@ 33.3\footnote{Therein, the ``dual feasible for $G$'' is defined under three conditions, one of which states that $B_{uv} = 0$ whenever the edge $\{i, j\}$ is \emph{not} in $G$. The correct condition, 
    however, is for when the edge \emph{is} in $G$.}),
    \begin{subequations} \label{eq:lovasz-number-definition}
    \begin{align}
        \vartheta(G) = &\max\nolimits_B \Lambda(B) \label{eq:lovasz-number-first}\\
                       &\text{ subject to:} \quad B \in \mathbb{R}^{n \times n} \text{ and } B \succeq 0 \\
                       &\quad B_{ii} = 1, \,\forall i \in \{1, \dots, n\} \\
                       &\quad B_{ij} = 0, \,\forall \{i, j\} \in G \label{eq:b-orthogonal-vectors} .
    \end{align}
    \end{subequations}
    This exactly matches the quantity in Eq.~\eqref{eq:max-eigenvalue-bound}, leading 
    to a generalisation of Eq.~(\ref{eq-fundamental}) as our main result:
    \begin{equation}
        \sum_{i=1}^n \ev{A_i}^2 \leq \lov .
        \label{eq:theta-bound}
    \end{equation}
    Note that these bounds are state independent; when the purity of the states is known, the Lovász number can also be employed to bound similar 
    expressions in various ways \cite{zhenpeng2022comm}.

    Also notably, there is a similar connection between linear expressions on squares of the expectation values, $\mathbb{E}_{\vec{w}} = \sum_{i=1}^n w_i \mean{A_i}^2$, where $\vec{w} = (w_1, \ldots, w_n)$, $w_i >0$, and the \emph{weighted Lovász number} $\vartheta(\overbar{G}, \vec{w})$. This is evidenced from the fact that, when considering $\mathbb{E}_{\vec{w}}$, Eq.~\eqref{eq:diagonals-mathcal-a-matrix} (and only that) changes to $\mathcal{A}_{ii} = w_i$. That is precisely how the representation of the Lovász number given in Eq.~\eqref{eq:lovasz-number-definition} changes in its weighted variation. Consequently,
    \begin{equation}
        \sum_{i=1}^n w_i \mean{A_i}^2 \leq \vartheta(\overbar{G}, \vec{w}).
        \label{eq:bound-with-weights}
    \end{equation}

    This allows us to extend our results for observables which are not necessarily dichotomic. Indeed, in this case, the elements $\mathcal{A}_{ii} = \langle A_i^2 \rangle_\rho$ are no longer equal to one but are upper bounded by $\Lambda(A_i^2)$. Then, by settings the weights to $\vec{w} = (\Lambda(A_1^2), \dots, \Lambda(A_n^2))$, we get that  
    $
    \sum_{i=1}^n \mean{A_i}^2 \leq \vartheta(\overbar{G}, \vec{w}).
    $
    The case for 
    dichotomic observables is recovered by unit weights.
    Computing $\vartheta(\overbar{G}, \vec{w})$ is also a small instance of a semidefinite program, thus the bound is easily computable \cite{knuth1994sandwich}.
    
    \section{Immediate consequences} When all the observables under consideration anticommute pairwise, the anticommutativity graph is a complete graph. In this case, in Eq.~(\ref{eq:lovasz-number-definition}), $B$ is the identity matrix. Then, $\lov = 1$,
    recovering the result derived in Refs.~\cite{toth2005entanglement,wehner2008higher}. Oppositely, the Lovász number of an empty graph is $n$, leading to the trivial bound
    $\sum_i \ev{A_i}^2 \leq n$.
    
    As stated in the introduction, the smallest number of pairwise anticommuting subsets of $\{ A_i \}$ is an upper bound for $\mathbb{E}$. In fact, this corresponds to the chromatic number of the associated graph $G$, which is always an upper bound on the Lovász number. Since there exist graphs for which $\lov < \chi(G)$, pairwise anticommutation relations alone do not always lead to the best upper bounds.
    
    In other words, our result shows that bounding $\mathbb{E}$ by the smallest number of pairwise anticommuting subsets, as it was usually done (e.g., Ref.\@ \cite{kurzynski2011correlation}), leads to loose bounds in all cases where the anticommutativity graph is not a perfect graph. For example, Fig.~\ref{fig:summary} shows a set of five observables that have a star anticommutativity graph. Grouping them in pairwise anticommuting subsets leads to an upper bound of $\chi(G) = 3$. On the other hand, the Lovász number of $\overbar{G}$ is equal to $\sqrt{5} \simeq 2.24$, from which we learn that the graph is not perfect and directly see that upper bounds on $\mathbb{E}$ cannot be deduced only from pairwise anticommutation relations. This is quite a significant gap, especially when one has quantum informational applications in mind, and in fact it can get much larger, since it is known that, for any $\varepsilon > 0$, there are graphs for which $\chi(G) > \lov n^{1-\varepsilon}$ \cite{feige1995randomized}.

\section{Including commutativity relations}
    If one considers tensor products of Pauli matrices, the observables $A_i$ have the 
    additional property that they can only commute or anticommute. In this case, one can 
    clearly achieve $\sum_i \ev{A_i}^2 = \omega(G)$ by choosing a common eigenstate for
    the largest commuting set. So, when $\omega(G) = \lov$, the bound in Eq.~(\ref{eq:theta-bound}) is tight. This happens for instance for all perfect graphs, but not only. 
    Interesting perfect graphs include all bipartite graphs, even cycles, and chordal 
    graphs. In fact, from the \emph{strong perfect graph theorem}, a graph is imperfect 
    if and only if it contains some odd cycle larger or equal than a pentagon (or the 
    complement of one) as an induced subgraph  \cite{chudnovsky2006strong}. Therefore, these 
    are the only cases where the bound in Eq.~\eqref{eq:theta-bound} may not be tight --- the 
    smallest example of such a possibility being the pentagon itself.

    However, for observables $\{A_i\}_{i=1}^5$ with a star anticommutativity graph such as in Fig.~\ref{fig:summary}, we can show that $\max_\varrho \sum_{i=1}^5 \mean{A_i}^2 = \omega(G) = 2$. Indeed, the product observables of the form $\{A_iA_{(i \mod 5) +1}\}_{i=1}^5$ have a fully connected anticommutativity graph, i.e., $\sum_{i=1}^5 \langle A_iA_{i+1} \rangle ^2 \leq 1$, which can be used to prove the claim. This naturally raises the question of whether the clique number of $G$ could be the actual tight bound, but a counterexample to this conjecture has been found \cite{counterexample}, showing that eigenstates of the largest commuting subset do not always maximise $\mathbb{E}$.

\section{Imprecise observables}
    As noted above, if no pair of observables in a given set $\{ A_i \}_{i=1}^n$ anticommute, then $\vartheta(\overbar{G}) = n$ is a trivial bound. This can be improved if we have access to more information regarding the observables. For example, one might wonder how the $\vartheta$-bound scales in the presence of a perturbation such that some pairs of observables ``almost'' anticommute. More than an interesting theoretical consideration, this is also in line with approaches to devise entanglement tests which are robust against imprecisely calibrated devices  \cite{moroder2012calibration,seevinck2007local,morelli2022entanglement}.

    Suppose, for that matter, that $\norm{ \{ A_i, A_j \}} < \eps_{ij}$ and define a critical $\eps$. We take norms that satisfy $\Lambda(\cdot) \leq \norm{\cdot}$, for instance, the operator norm. We say that any pair of measurements with $\eps_{ij} < \eps$ \emph{almost anticommute}. We now want to know how this affects our bound, as it for now solely depends on vanishing anticommutators. For that purpose, we build the matrix $\mathcal{E}(\vec{w})$ with elements $\sqrt{w_i w_j} \eps_{ij}$ if $A_i$ and $A_j$ almost anticommute and zero otherwise. We can then show that (see Appendix)
    \begin{equation}\label{eq:epsilon-bound}
        \sum_i w_i \mean{A_i}^2 \leq \vartheta(\overbar{G}_\eps, \vec{w}) + \frac{1}{2} \Lambda\left[\mathcal{E}(\vec{w})\right],
    \end{equation}
    where two vertices in $\overbar{G}_\eps$ are connected if their corresponding observables almost anticommute.
    Notice that $\Lambda\left(\mathcal{E}(\vec{w})\right) \leq \eps \Lambda(\overbar{G}_\eps)$, where by $\Lambda(\overbar{G}_\eps)$ we understand the largest eigenvalue of the adjacency matrix of $\overbar{G}_\eps$.

    Equation~\eqref{eq:epsilon-bound} is surprisingly simple. Moreover, as will be discussed later, the fact that it decouples the contributions of the idealised observables and the perturbation also greatly simplifies the design and analysis of entanglement tests under more realistic conditions.
    
\section{Application: Entanglement witnesses}
    Entanglement witnesses are the main tool for entanglement detection in practice.
    However, constructing experimentally friendly witnesses can be difficult, since obtaining the bounds frequently require them to have a particular structure, which does not necessarily match what can be implemented in a laboratory. Furthermore, it has been noticed that even small imprecisions in the measurements can lead to false positives in entanglement and nonlocality detection \cite{moroder2012calibration,seevinck2007local, morelli2022entanglement}, but this analysis could only be done for the simplest examples. Here, we demonstrate how Eq.~\eqref{eq:epsilon-bound} gives a straightforward solution to both these problems.

    Recall that a two-party state $\varrho$ is separable ($\varrho \in$ \textsc{sep}) 
   if it can be written as a convex combination of product states, $\varrho = \sum_k 
   p_k \varrho_k^{A} \otimes \varrho_k^{B}$. Rewriting Eq.\@ \eqref{eq:theta-bound} 
   and making use of the the Cauchy–Schwarz inequality, we get 
    \begin{equation} \label{eq:sepcrit}
        \max_{\varrho \in \text{SEP}} \sum_{i=1}^n \mean{A_i\otimes B_i} \leq \sqrt{\vartheta(\overbar{G}_A)\vartheta(\overbar{G}_B)} =: \vartheta_{AB} ,
    \end{equation}
    where $\vartheta(\overbar{G}_A)$ is the Lovász number of the anticommutativity graph of $\{A_i\}$ and similarly for $\vartheta(\overbar{G}_B)$. 
    This separability criterion naturally leads to entanglement witnesses of the form \cite{guhne2009entanglement}
    \begin{equation} \label{eq:witness}
        W_\mathbb{E} = \vartheta_{AB} \id - \sum_i A_i \otimes B_i.
    \end{equation}
   These operators are required to have $\tr{W_\mathbb{E} \varrho} \geq 0$ for separable $\varrho$, which directly follows, and to have $\tr{W_\mathbb{E} \varrho} < 0$ for at least one entangled state, which is fulfilled when the observables are chosen insightfully. For cases where the $\vartheta(\overbar{G}_A)$ and $\vartheta(\overbar{G}_B)$ bounds are tight, $W_\mathbb{E}$ is a weakly optimal entanglement witness, i.e.,  there exists at least one separable state $\varrho_s$ such that $\tr{W \varrho_s}=0$.
    
    For concreteness, let us choose an instance of a \textsc{ppt} entangled state,
    \begin{equation}
        \varrho_I = \frac{1}{6} \left( \varrho_{\id Y} + \varrho_{XX} + \varrho_{YZ} + \varrho_{ZX} + \varrho_{ZY} + \varrho_{ZZ} \right),
    \end{equation}
    where $\varrho_{AB} $ is the projector onto the pure state $\ket{\psi_{AB}} = \nicefrac{1}{2} \left( \id \id A B \sum_{i=0}^3 \ket{i} \ket{i} \right)$, $A,B \in \{\id, X, Y, Z\}$ \cite{benatti2004non}. Physically, this state can be seen as a
    mixture of six pairs of Bell states ($\ket{\beta_i}$) of the form $\ketbra{\beta_i}{\beta_i}_{A' B'} \otimes \ketbra{\beta_j}{\beta_j}_{A'' B''}$,
    where one party holds the qubits $A'$ and $A''$, and the other $B'$ and $B''$ \cite{benatti2004non, moroder2012calibration}.
    As sets of observables, we take the elements in the Bloch decomposition of $\varrho_I$ which correspond to non-zero coefficients, that is,
    \begin{align} \label{eq:setofobsexample}
    \{A_i/ & B_i\} 
     = 
    \{ \pm \id X, \id Y, \pm \id Z, \pm X \id , XX, \pm XY , \pm XZ, \nonumber 
    \\ 
    & Y \id, Y X, YY, \pm YZ, Z \id, \pm ZX, \pm ZY, \pm ZZ\},     
    \end{align}
    where the upper signs are for $\{A_i\}$.
    These sets have the same anticommutativity graph, for which 
    $ \lov = \omega(G) =3$. The witness based on these observables 
    detects $\nu \varrho_I + (1-\nu) \nicefrac{\id}{16}$ to 
    be entangled for $\nu > \nicefrac{3}{5}$, which is exactly 
    the separability bound \cite{moroder2012calibration}. Notice that for these sets of observables, the chromatic number $\chi(G)$ is equal to four. If the witness were constructed with this quantity instead of the Lovász number bound, it would only be able to detect entanglement for a visibility $\nu$ above $\nicefrac{4}{5}$. This illustrates how only considering sets of pairwise anticommuting observables can lead to untight bounds. 

    Now let us consider a more realistic scenario and analyse how imprecisions in the measurements affect the entanglement detection threshold. To this end, let $\{ A_i \}$ be the target observables and $\{ \tilde{A}_i \}$ the idealised ones. We choose an imprecision bound $\varepsilon$ such that the almost anticommutativity graph $\overbar{G}_\eps$ is the same as $\overbar{G}$, and for simplicity we consider that all $\eps_{ij}$ are equal to $\eps$. Equation \eqref{eq:epsilon-bound} then comes in handy, as it enables us to construct the witness $\left(3+4\eps\right) \id - \sum_i \tilde{A}_i \otimes \tilde{B}_i$, which can detect entangled states when $\eps < \nicefrac{1}{2}$. For our \textsc{ppt} entangled state, the witness can certify that $\nu \varrho_I + (1-\nu) \nicefrac{\id}{16}$ is entangled for $\nu > \nicefrac{(3+4\eps)}{5}$. This exemplifies that, although Eq.\@ \eqref{eq:witness} is a witness, if it is implemented by measuring imprecisely the observables of Eq.\@ \eqref{eq:setofobsexample}, a negative expectation value can no longer be taken as a certification that the state is entangled.  

\section{Application: Uncertainty relations}
    As explained already in the introduction, our result in Eq.~\eqref{eq:theta-bound} gives rise to the uncertainty
    relation
    \begin{equation} \label{eq:metaur}
        \sum_{i=1}^n \Delta^2(A_i) \geq n - \lov.
    \end{equation}
    Using the approach of local uncertainty relations 
    \cite{hofmann2003violation} this can be turned into a nonlinear entanglement
    criterion: Suppose that a local uncertainty relation $\sum_{i=1}^n 
    \Delta^2(A_i)_{\varrho_A} \geq C_A$ is true for all $d_A$-dimensional 
    quantum states $\varrho_A$ and $\sum_{i=1}^n \Delta^2(B_i)_{\varrho_B} 
    \geq C_B$, similarly, for any state of dimension $d_B$. Then, if we 
    define on the bipartite system
    $M_i \equiv  A_i \otimes \id - \id \otimes B_i$, for separable states, $\sum_{i=1}^n \Delta^2(M_i)_{\varrho} \geq C_A + C_B$ holds.
    Combining this method with Eq.~(\ref{eq:metaur})  leads to 
    nonlinear witnesses of the form
    \begin{align} 
    &\frac{1}{2} \big[\vartheta(\overbar{G}_A) + \vartheta(\overbar{G}_B)\big]
    - \sum_{i=1}^n \mean{ A_i \otimes B_i } 
    \nonumber \\
    & \quad -  \frac{1}{2}
    \sum_{i=1}^n \mean{ A_i \otimes \id - \id \otimes B_i }^2 \underset{\textsc{sep}}{\geq} 0 ,
     \label{eq:nonlinwitness}
    \end{align}
    where $\overbar{G}_A$ and $\overbar{G}_B$ are the anticommutativity 
    graphs corresponding to $\{ A_i \}$ and $\{ B_i \}$, respectively. 
    If $\vartheta(\overbar{G}_A) = \vartheta(\overbar{G}_B) $ this 
    criterion is obviously stronger than Eq.~(\ref{eq:witness}), see 
    also Ref.~\cite{guehne2006entanglement}.
    
    From the variance uncertainty relation in Eq.~(\ref{eq:metaur}), it is also 
    possible to derive several entropic uncertainty relations. The latter are sometimes considered to be a more fundamental formulation 
    of uncertainty relations, and can be used in security proofs for quantum key 
    distribution and as entanglement witnesses, among other applications \cite{coles2017entropic}.

    As an example, when $\lov$ is an integer (e.g.,\@ for perfect graphs), 
    Theorem 6 in Ref.\@ \cite{niekamp2012entropic} is straightforwardly 
    generalisable to our case and leads to the tight bound
    \begin{equation}
        \sum_{i=1}^n S(A_i|\varrho) \geq \left[ n-\lov \right] S_0.
        \label{eq:entropic-bounds}
    \end{equation}
    Here, $S$ is any entropy that is concave in $\mean{A_i}^2$,\footnote{Concavity in $\mean{A_i}^2$ here means that the function $\tilde{S}(x) := S(A_i|\varrho) = 
    S\left(P_\pm \right)$ is concave in $x=\mean{A_i}^2$, where 
    $P_\pm = (\nicefrac{1 \pm \sqrt{x}}{2}, \nicefrac{1 \mp \sqrt{x}}{2})$ are 
    the two possible probability distribution arising from measuring $A_i$.} $S(A_i|\varrho)$ is the entropy of the probability distribution gotten 
    from measuring $A_i$ on $\varrho$ and $S_0$ is the entropy value of the 
    flat probability distribution. The proof of the statement proceeds just as
    in Ref.~\cite{niekamp2012entropic}, with the difference that the allowed
    domain of values for $\mean{A_1}^2, \dots , \mean{A_n}^2,$ has now as
    extremal points all vectors with up to $\lov$ entries ``1'' and zeroes
    elsewhere. Prominent examples of entropies that have the desired concavity 
    property, are the Shannon entropy and the Tsallis entropies $S_q^T$ for 
    $1<q<2$ and $3<q$ \cite{niekamp2012entropic}.
    
    Interestingly, Eq.~\eqref{eq:entropic-bounds} can easily be turned into 
    entropic criteria for steering and entanglement. To understand how, suppose 
    that $\sum_{i=1}^n S(A_i \vert \varrho_A) \geq C$ holds for all $d_A$-dimensional quantum states $\varrho_A$. Then, for all separable bipartite $d_A d_B$-dimensional states and  any set of $n$ observables $\{B_i\}$,
    $\min_{\varrho \in \textsc{sep}} \sum_{i=1}^n S(A_i \otimes B_i \vert \varrho) \geq C$ \cite{guhne2004entropic}. Even more strongly, all probability distributions arising from measuring any unsteerable state fulfil the same condition, thus these same expressions can also be used as entropic steering criteria \cite{costa2018steering}.

    As a closing comment we recall that, in the two-dimensional case, 
    $\Delta^2(X) + \Delta^2(Y) + \Delta^2(Z) \geq 2$ completely characterises 
    the set of states, in the sense that any probability distribution obeying 
    it can originate from a valid quantum state. This also holds for the 
    extension to larger sets of pairwise anticommuting sets of observables but is, however, not true in general for the constraints in Eq.~\eqref{eq:metaur}. 
    To build an example, start from the Pauli basis in dimension $d = 4$, given by $\{ A_i\} = \{ \id, X, Y, Z \}^{\otimes 2}$. In this case, $\lov = 4$, and from Eq.~\eqref{eq:theta-bound}, $\sum_{i=1}^{16} \mean{A_i}_\varrho^2 \leq 4$, 
    implying that the length of the Bloch vector of any ququart is bounded by $4$. 
    But it is well-known that there are vectors of length $4$ which do not 
    represent quantum states \cite{kimura2003bloch,byrd2003characterization}, 
    therefore there must be higher-order constraints not encoded in 
    Eq.~\eqref{eq:metaur}. Finding such relations is an interesting topic for further
    research.
    
\section{Discussion}
    Through exploring a link between observables and graph theory, we have proven 
    that the sum of squares of expectation values of any set of 
    observables is bounded by the Lovász number of their anticommutativity 
    graph. This is a significant improvement to the previously known bound based on the chromatic number.
    Our result found further applications for the characterisation of 
    entanglement and quantum steering, and also in various uncertainty 
    relations. Notably, we were able to build witnesses robust to imprecise observables.  
    
     Regarding extensions of our results, it could also be insightful to consider other equivalent definitions of the Lovász number. Our choice, presented in Eq.~\eqref{eq:lovasz-number-definition}, was based on the straightforward connection with Eq.~\eqref{eq:max-eigenvalue-bound}. However, others may provide further physical intuition. Of particular interest is the representation given in Theorem 5 of \cite{lovasz1979shannon}, which can be more closely related to quantum structures, such as state vectors and measurements. Indeed, it was already shown to be connected to quantum bounds in contextuality inequalities \cite{cabello2014graph}. Similarly, it could also be interesting to find closer correspondences with other approaches to similar problems, such as the one via joint numerical ranges (e.g., \cite{schwonnek2017state,konrad2019geometric}).
    
    Applications-wise, the entanglement witnesses we have shown are just simple examples, and much in the way of tailoring them to particular families of states can be done. Apart from standard entanglement theory, bounds for special cases of $\mathbb{E}$ were previously applied to the network entanglement compatibility problem \cite{hansenne2022symmetries}, and we believe our more general bounds can be further employed in network scenarios and lead to stronger certificates. Moreover, our results imply bounds on the trace of covariance matrices, which can be used as network entanglement criteria as well. 

    \textit{Note added.} During the course of publication we became aware that a result analogous to Eq.~\eqref{eq:theta-bound} was found in the context of fermionic Hamiltonian optimisation \cite{hastings2022optimizing}. 

\section*{Acknowledgments}
    Many thanks to J.\@ Carrasco, D.\@ García-Martín, F.\@ Huber, B.\@ Kraus and A.\@ Perez-Salinas (among many other participants of the ``Entanglement in Action'' conference in Benasque), and also to J.\@ L.\@ Bönsel, M.\@ Terra Cunha, L.\@ Ligthart, H.\@ C.\@ Nguyen, T.\@ Ohst, M.\@ Plávala, R.\@ Schwonnek, L.\@ Vandré, R.\@ Werner, Z.-P.\@ Xu, and B.\@ Yadin for interesting discussions. 
    This work was financially supported by the Deutsche Forschungsgemeinschaft (DFG, German Research Foundation, project numbers 447948357 and 440958198), the Sino-German Center for Research Promotion (Project M-0294), and the House of Young Talents of the University of Siegen.

\section*{Appendix}
    In this Appendix, we show how to derive the upper bound in Eq.\@ \eqref{eq:epsilon-bound}. In the same way as the proof of the main result, we can show that $\norm{\vec{a}}^2 \sum_i w_i \mean{A_i}^2 \leq \sum_{ij} a_i a_j \sqrt{w_i w_j} \mean{A_iA_j}$, with $a_i = \sqrt{w_i} \mean{A_i}$. Then, we split the right hand side into a sum over the pairs $ij$ that do not almost anticommute and a sum over the pair that do. The former sum is upperbounded by $\norm{\vec{a}}^2 \vartheta(\overbar{G}_\eps, \vec{w})$ as before, whereas the latter sum is upperbounded by $\frac{\norm{\vec{a}}^2}{2} \Lambda(\mathcal{E}(\vec{w}))$, which finishes the proof.

\bibliographystyle{apsrev4-1}
\bibliography{bibliography}

\begin{thebibliography}{48}%
\makeatletter
\providecommand \@ifxundefined [1]{%
 \@ifx{#1\undefined}
}%
\providecommand \@ifnum [1]{%
 \ifnum #1\expandafter \@firstoftwo
 \else \expandafter \@secondoftwo
 \fi
}%
\providecommand \@ifx [1]{%
 \ifx #1\expandafter \@firstoftwo
 \else \expandafter \@secondoftwo
 \fi
}%
\providecommand \natexlab [1]{#1}%
\providecommand \enquote  [1]{``#1''}%
\providecommand \bibnamefont  [1]{#1}%
\providecommand \bibfnamefont [1]{#1}%
\providecommand \citenamefont [1]{#1}%
\providecommand \href@noop [0]{\@secondoftwo}%
\providecommand \href [0]{\begingroup \@sanitize@url \@href}%
\providecommand \@href[1]{\@@startlink{#1}\@@href}%
\providecommand \@@href[1]{\endgroup#1\@@endlink}%
\providecommand \@sanitize@url [0]{\catcode `\\12\catcode `\$12\catcode
  `\&12\catcode `\#12\catcode `\^12\catcode `\_12\catcode `\%12\relax}%
\providecommand \@@startlink[1]{}%
\providecommand \@@endlink[0]{}%
\providecommand \url  [0]{\begingroup\@sanitize@url \@url }%
\providecommand \@url [1]{\endgroup\@href {#1}{\urlprefix }}%
\providecommand \urlprefix  [0]{URL }%
\providecommand \Eprint [0]{\href }%
\providecommand \doibase [0]{http://dx.doi.org/}%
\providecommand \selectlanguage [0]{\@gobble}%
\providecommand \bibinfo  [0]{\@secondoftwo}%
\providecommand \bibfield  [0]{\@secondoftwo}%
\providecommand \translation [1]{[#1]}%
\providecommand \BibitemOpen [0]{}%
\providecommand \bibitemStop [0]{}%
\providecommand \bibitemNoStop [0]{.\EOS\space}%
\providecommand \EOS [0]{\spacefactor3000\relax}%
\providecommand \BibitemShut  [1]{\csname bibitem#1\endcsname}%
\let\auto@bib@innerbib\@empty
\bibitem [{\citenamefont {Heisenberg}(1927)}]{Heisenberg_1927}%
  \BibitemOpen
  \bibfield  {author} {\bibinfo {author} {\bibfnamefont {W.}~\bibnamefont
  {Heisenberg}},\ }\href {\doibase 10.1007/bf01397280} {\bibfield  {journal}
  {\bibinfo  {journal} {Z. Physik}\ }\textbf {\bibinfo {volume} {43}},\
  \bibinfo {pages} {172} (\bibinfo {year} {1927})}\BibitemShut {NoStop}%
\bibitem [{\citenamefont {Kennard}(1927)}]{Kennard_1927}%
  \BibitemOpen
  \bibfield  {author} {\bibinfo {author} {\bibfnamefont {E.~H.}\ \bibnamefont
  {Kennard}},\ }\href {\doibase 10.1007/BF01391200} {\bibfield  {journal}
  {\bibinfo  {journal} {Z. Physik}\ }\textbf {\bibinfo {volume} {44}},\
  \bibinfo {pages} {326} (\bibinfo {year} {1927})}\BibitemShut {NoStop}%
\bibitem [{\citenamefont {Robertson}(1929)}]{robertson29}%
  \BibitemOpen
  \bibfield  {author} {\bibinfo {author} {\bibfnamefont {H.~P.}\ \bibnamefont
  {Robertson}},\ }\href {\doibase 10.1103/PhysRev.34.163} {\bibfield  {journal}
  {\bibinfo  {journal} {Phys. Rev.}\ }\textbf {\bibinfo {volume} {34}},\
  \bibinfo {pages} {163} (\bibinfo {year} {1929})}\BibitemShut {NoStop}%
\bibitem [{\citenamefont {Robertson}(1934)}]{robertson34}%
  \BibitemOpen
  \bibfield  {author} {\bibinfo {author} {\bibfnamefont {H.~P.}\ \bibnamefont
  {Robertson}},\ }\href {\doibase 10.1103/PhysRev.46.794} {\bibfield  {journal}
  {\bibinfo  {journal} {Phys. Rev.}\ }\textbf {\bibinfo {volume} {46}},\
  \bibinfo {pages} {794} (\bibinfo {year} {1934})}\BibitemShut {NoStop}%
\bibitem [{\citenamefont {Wehner}\ and\ \citenamefont
  {Winter}(2010)}]{wehner2010entropic}%
  \BibitemOpen
  \bibfield  {author} {\bibinfo {author} {\bibfnamefont {S.}~\bibnamefont
  {Wehner}}\ and\ \bibinfo {author} {\bibfnamefont {A.}~\bibnamefont
  {Winter}},\ }\href
  {https://iopscience.iop.org/article/10.1088/1367-2630/12/2/025009/meta}
  {\bibfield  {journal} {\bibinfo  {journal} {New J. Phys.}\ }\textbf {\bibinfo
  {volume} {12}},\ \bibinfo {pages} {025009} (\bibinfo {year}
  {2010})}\BibitemShut {NoStop}%
\bibitem [{\citenamefont {Coles}\ \emph {et~al.}(2017)\citenamefont {Coles},
  \citenamefont {Berta}, \citenamefont {Tomamichel},\ and\ \citenamefont
  {Wehner}}]{coles2017entropic}%
  \BibitemOpen
  \bibfield  {author} {\bibinfo {author} {\bibfnamefont {P.~J.}\ \bibnamefont
  {Coles}}, \bibinfo {author} {\bibfnamefont {M.}~\bibnamefont {Berta}},
  \bibinfo {author} {\bibfnamefont {M.}~\bibnamefont {Tomamichel}}, \ and\
  \bibinfo {author} {\bibfnamefont {S.}~\bibnamefont {Wehner}},\ }\href
  {https://journals.aps.org/rmp/abstract/10.1103/RevModPhys.89.015002}
  {\bibfield  {journal} {\bibinfo  {journal} {Rev. Mod. Phys.}\ }\textbf
  {\bibinfo {volume} {89}},\ \bibinfo {pages} {015002} (\bibinfo {year}
  {2017})}\BibitemShut {NoStop}%
\bibitem [{\citenamefont {Busch}\ \emph
  {et~al.}(2014{\natexlab{a}})\citenamefont {Busch}, \citenamefont {Lahti},\
  and\ \citenamefont {Werner}}]{busch2014colloquium}%
  \BibitemOpen
  \bibfield  {author} {\bibinfo {author} {\bibfnamefont {P.}~\bibnamefont
  {Busch}}, \bibinfo {author} {\bibfnamefont {P.}~\bibnamefont {Lahti}}, \ and\
  \bibinfo {author} {\bibfnamefont {R.~F.}\ \bibnamefont {Werner}},\ }\href
  {\doibase 10.1103/RevModPhys.86.1261} {\bibfield  {journal} {\bibinfo
  {journal} {Rev. Mod. Phys.}\ }\textbf {\bibinfo {volume} {86}},\ \bibinfo
  {pages} {1261} (\bibinfo {year} {2014}{\natexlab{a}})}\BibitemShut {NoStop}%
\bibitem [{\citenamefont {Schwonnek}(2018)}]{schwonnek2018phd}%
  \BibitemOpen
  \bibfield  {author} {\bibinfo {author} {\bibfnamefont {R.}~\bibnamefont
  {Schwonnek}},\ }\emph {\bibinfo {title} {Uncertainty relations in quantum
  theory}},\ \href {https://doi.org/10.15488/3600} {Ph.D. thesis},\ \bibinfo
  {school} {Gottfried Wilhelm Leibniz Universität Hannover} (\bibinfo {year}
  {2018})\BibitemShut {NoStop}%
\bibitem [{\citenamefont {T\'oth}\ and\ \citenamefont
  {G\"uhne}(2005)}]{toth2005entanglement}%
  \BibitemOpen
  \bibfield  {author} {\bibinfo {author} {\bibfnamefont {G.}~\bibnamefont
  {T\'oth}}\ and\ \bibinfo {author} {\bibfnamefont {O.}~\bibnamefont
  {G\"uhne}},\ }\href {\doibase 10.1103/PhysRevA.72.022340} {\bibfield
  {journal} {\bibinfo  {journal} {Phys. Rev. A}\ }\textbf {\bibinfo {volume}
  {72}},\ \bibinfo {pages} {022340} (\bibinfo {year} {2005})}\BibitemShut
  {NoStop}%
\bibitem [{\citenamefont {Szyma{\'n}ski}\ \emph {et~al.}(2018)\citenamefont
  {Szyma{\'n}ski}, \citenamefont {Weis},\ and\ \citenamefont
  {{\.Z}yczkowski}}]{szymanski2018classification}%
  \BibitemOpen
  \bibfield  {author} {\bibinfo {author} {\bibfnamefont {K.}~\bibnamefont
  {Szyma{\'n}ski}}, \bibinfo {author} {\bibfnamefont {S.}~\bibnamefont {Weis}},
  \ and\ \bibinfo {author} {\bibfnamefont {K.}~\bibnamefont {{\.Z}yczkowski}},\
  }\href {https://doi.org/10.48550/arXiv.1603.06569} {\bibfield  {journal}
  {\bibinfo  {journal} {Linear Algebra Appl.}\ }\textbf {\bibinfo {volume}
  {545}},\ \bibinfo {pages} {148} (\bibinfo {year} {2018})}\BibitemShut
  {NoStop}%
\bibitem [{\citenamefont {Schulte-Herbr{\"u}ggen}\ \emph
  {et~al.}(2008)\citenamefont {Schulte-Herbr{\"u}ggen}, \citenamefont {Dirr},
  \citenamefont {Helmke},\ and\ \citenamefont
  {Glaser}}]{schulte2008significance}%
  \BibitemOpen
  \bibfield  {author} {\bibinfo {author} {\bibfnamefont {T.}~\bibnamefont
  {Schulte-Herbr{\"u}ggen}}, \bibinfo {author} {\bibfnamefont {G.}~\bibnamefont
  {Dirr}}, \bibinfo {author} {\bibfnamefont {U.}~\bibnamefont {Helmke}}, \ and\
  \bibinfo {author} {\bibfnamefont {S.~J.}\ \bibnamefont {Glaser}},\ }\href
  {https://doi.org/10.1080/03081080701544114} {\bibfield  {journal} {\bibinfo
  {journal} {Lin. Multilin. Alg.}\ }\textbf {\bibinfo {volume} {56}},\ \bibinfo
  {pages} {3} (\bibinfo {year} {2008})}\BibitemShut {NoStop}%
\bibitem [{\citenamefont {Gawron}\ \emph {et~al.}(2010)\citenamefont {Gawron},
  \citenamefont {Pucha{\l}a}, \citenamefont {Miszczak}, \citenamefont
  {Skowronek},\ and\ \citenamefont {{\.Z}yczkowski}}]{gawron2010restricted}%
  \BibitemOpen
  \bibfield  {author} {\bibinfo {author} {\bibfnamefont {P.}~\bibnamefont
  {Gawron}}, \bibinfo {author} {\bibfnamefont {Z.}~\bibnamefont {Pucha{\l}a}},
  \bibinfo {author} {\bibfnamefont {J.~A.}\ \bibnamefont {Miszczak}}, \bibinfo
  {author} {\bibfnamefont {{\L}.}~\bibnamefont {Skowronek}}, \ and\ \bibinfo
  {author} {\bibfnamefont {K.}~\bibnamefont {{\.Z}yczkowski}},\ }\href
  {https://aip.scitation.org/doi/full/10.1063/1.3496901} {\bibfield  {journal}
  {\bibinfo  {journal} {J. Math. Phys.}\ }\textbf {\bibinfo {volume} {51}},\
  \bibinfo {pages} {102204} (\bibinfo {year} {2010})}\BibitemShut {NoStop}%
\bibitem [{\citenamefont {Wehner}\ and\ \citenamefont
  {Winter}(2008)}]{wehner2008higher}%
  \BibitemOpen
  \bibfield  {author} {\bibinfo {author} {\bibfnamefont {S.}~\bibnamefont
  {Wehner}}\ and\ \bibinfo {author} {\bibfnamefont {A.}~\bibnamefont
  {Winter}},\ }\href {https://aip.scitation.org/doi/full/10.1063/1.2943685}
  {\bibfield  {journal} {\bibinfo  {journal} {J. Math. Phys.}\ }\textbf
  {\bibinfo {volume} {49}},\ \bibinfo {pages} {062105} (\bibinfo {year}
  {2008})}\BibitemShut {NoStop}%
\bibitem [{\citenamefont {Pauli}(1927)}]{pauli1927matrices}%
  \BibitemOpen
  \bibfield  {author} {\bibinfo {author} {\bibfnamefont {W.}~\bibnamefont
  {Pauli}},\ }\href {\doibase 10.1007/BF01397326} {\bibfield  {journal}
  {\bibinfo  {journal} {Z. Physik}\ }\textbf {\bibinfo {volume} {43}},\
  \bibinfo {pages} {601} (\bibinfo {year} {1927})}\BibitemShut {NoStop}%
\bibitem [{\citenamefont {Feynman}\ \emph {et~al.}(1957)\citenamefont
  {Feynman}, \citenamefont {Vernon~Jr},\ and\ \citenamefont
  {Hellwarth}}]{feynman1957geometrical}%
  \BibitemOpen
  \bibfield  {author} {\bibinfo {author} {\bibfnamefont {R.~P.}\ \bibnamefont
  {Feynman}}, \bibinfo {author} {\bibfnamefont {F.~L.}\ \bibnamefont
  {Vernon~Jr}}, \ and\ \bibinfo {author} {\bibfnamefont {R.~W.}\ \bibnamefont
  {Hellwarth}},\ }\href {https://doi.org/10.1063/1.1722572} {\bibfield
  {journal} {\bibinfo  {journal} {J. Appl. Phys.}\ }\textbf {\bibinfo {volume}
  {28}},\ \bibinfo {pages} {49} (\bibinfo {year} {1957})}\BibitemShut {NoStop}%
\bibitem [{\citenamefont {Hofmann}\ and\ \citenamefont
  {Takeuchi}(2003)}]{hofmann2003violation}%
  \BibitemOpen
  \bibfield  {author} {\bibinfo {author} {\bibfnamefont {H.~F.}\ \bibnamefont
  {Hofmann}}\ and\ \bibinfo {author} {\bibfnamefont {S.}~\bibnamefont
  {Takeuchi}},\ }\href
  {https://journals.aps.org/pra/abstract/10.1103/PhysRevA.68.032103} {\bibfield
   {journal} {\bibinfo  {journal} {Phys. Rev. A}\ }\textbf {\bibinfo {volume}
  {68}},\ \bibinfo {pages} {032103} (\bibinfo {year} {2003})}\BibitemShut
  {NoStop}%
\bibitem [{\citenamefont {Portmann}\ and\ \citenamefont
  {Renner}(2022)}]{portmann2022security}%
  \BibitemOpen
  \bibfield  {author} {\bibinfo {author} {\bibfnamefont {C.}~\bibnamefont
  {Portmann}}\ and\ \bibinfo {author} {\bibfnamefont {R.}~\bibnamefont
  {Renner}},\ }\href {\doibase 10.1103/RevModPhys.94.025008} {\bibfield
  {journal} {\bibinfo  {journal} {Rev. Mod. Phys.}\ }\textbf {\bibinfo {volume}
  {94}},\ \bibinfo {pages} {025008} (\bibinfo {year} {2022})}\BibitemShut
  {NoStop}%
\bibitem [{\citenamefont {Niekamp}\ \emph {et~al.}(2012)\citenamefont
  {Niekamp}, \citenamefont {Kleinmann},\ and\ \citenamefont
  {G{\"u}hne}}]{niekamp2012entropic}%
  \BibitemOpen
  \bibfield  {author} {\bibinfo {author} {\bibfnamefont {S.}~\bibnamefont
  {Niekamp}}, \bibinfo {author} {\bibfnamefont {M.}~\bibnamefont {Kleinmann}},
  \ and\ \bibinfo {author} {\bibfnamefont {O.}~\bibnamefont {G{\"u}hne}},\
  }\href {https://aip.scitation.org/doi/full/10.1063/1.3678200} {\bibfield
  {journal} {\bibinfo  {journal} {J. Math. Phys.}\ }\textbf {\bibinfo {volume}
  {53}},\ \bibinfo {pages} {012202} (\bibinfo {year} {2012})}\BibitemShut
  {NoStop}%
\bibitem [{\citenamefont {Kurzy\ifmmode~\acute{n}\else \'{n}\fi{}ski}\ \emph
  {et~al.}(2011)\citenamefont {Kurzy\ifmmode~\acute{n}\else \'{n}\fi{}ski},
  \citenamefont {Paterek}, \citenamefont {Ramanathan}, \citenamefont
  {Laskowski},\ and\ \citenamefont {Kaszlikowski}}]{kurzynski2011correlation}%
  \BibitemOpen
  \bibfield  {author} {\bibinfo {author} {\bibfnamefont {P.}~\bibnamefont
  {Kurzy\ifmmode~\acute{n}\else \'{n}\fi{}ski}}, \bibinfo {author}
  {\bibfnamefont {T.}~\bibnamefont {Paterek}}, \bibinfo {author} {\bibfnamefont
  {R.}~\bibnamefont {Ramanathan}}, \bibinfo {author} {\bibfnamefont
  {W.}~\bibnamefont {Laskowski}}, \ and\ \bibinfo {author} {\bibfnamefont
  {D.}~\bibnamefont {Kaszlikowski}},\ }\href {\doibase
  10.1103/PhysRevLett.106.180402} {\bibfield  {journal} {\bibinfo  {journal}
  {Phys. Rev. Lett.}\ }\textbf {\bibinfo {volume} {106}},\ \bibinfo {pages}
  {180402} (\bibinfo {year} {2011})}\BibitemShut {NoStop}%
\bibitem [{\citenamefont {Hansenne}\ \emph {et~al.}(2022)\citenamefont
  {Hansenne}, \citenamefont {Xu}, \citenamefont {Kraft},\ and\ \citenamefont
  {G{\"u}hne}}]{hansenne2022symmetries}%
  \BibitemOpen
  \bibfield  {author} {\bibinfo {author} {\bibfnamefont {K.}~\bibnamefont
  {Hansenne}}, \bibinfo {author} {\bibfnamefont {Z.-P.}\ \bibnamefont {Xu}},
  \bibinfo {author} {\bibfnamefont {T.}~\bibnamefont {Kraft}}, \ and\ \bibinfo
  {author} {\bibfnamefont {O.}~\bibnamefont {G{\"u}hne}},\ }\href
  {https://www.nature.com/articles/s41467-022-28006-3} {\bibfield  {journal}
  {\bibinfo  {journal} {Nat. Commun.}\ }\textbf {\bibinfo {volume} {13}},\
  \bibinfo {pages} {1} (\bibinfo {year} {2022})}\BibitemShut {NoStop}%
\bibitem [{\citenamefont {Ozawa}(2003)}]{ozawa2003universally}%
  \BibitemOpen
  \bibfield  {author} {\bibinfo {author} {\bibfnamefont {M.}~\bibnamefont
  {Ozawa}},\ }\href {\doibase 10.1103/PhysRevA.67.042105} {\bibfield  {journal}
  {\bibinfo  {journal} {Phys. Rev. A}\ }\textbf {\bibinfo {volume} {67}},\
  \bibinfo {pages} {042105} (\bibinfo {year} {2003})}\BibitemShut {NoStop}%
\bibitem [{\citenamefont {Busch}\ \emph
  {et~al.}(2014{\natexlab{b}})\citenamefont {Busch}, \citenamefont {Lahti},\
  and\ \citenamefont {Werner}}]{busch2014heisenberg}%
  \BibitemOpen
  \bibfield  {author} {\bibinfo {author} {\bibfnamefont {P.}~\bibnamefont
  {Busch}}, \bibinfo {author} {\bibfnamefont {P.}~\bibnamefont {Lahti}}, \ and\
  \bibinfo {author} {\bibfnamefont {R.~F.}\ \bibnamefont {Werner}},\ }\href
  {\doibase 10.1103/PhysRevA.89.012129} {\bibfield  {journal} {\bibinfo
  {journal} {Phys. Rev. A}\ }\textbf {\bibinfo {volume} {89}},\ \bibinfo
  {pages} {012129} (\bibinfo {year} {2014}{\natexlab{b}})}\BibitemShut
  {NoStop}%
\bibitem [{\citenamefont {Maccone}\ and\ \citenamefont
  {Pati}(2014)}]{maccone2014stronger}%
  \BibitemOpen
  \bibfield  {author} {\bibinfo {author} {\bibfnamefont {L.}~\bibnamefont
  {Maccone}}\ and\ \bibinfo {author} {\bibfnamefont {A.~K.}\ \bibnamefont
  {Pati}},\ }\href {\doibase 10.1103/PhysRevLett.113.260401} {\bibfield
  {journal} {\bibinfo  {journal} {Phys. Rev. Lett.}\ }\textbf {\bibinfo
  {volume} {113}},\ \bibinfo {pages} {260401} (\bibinfo {year}
  {2014})}\BibitemShut {NoStop}%
\bibitem [{\citenamefont {Szyma{\'{n}}ski}\ and\ \citenamefont
  {{\.{Z}}yczkowski}(2019)}]{konrad2019geometric}%
  \BibitemOpen
  \bibfield  {author} {\bibinfo {author} {\bibfnamefont {K.}~\bibnamefont
  {Szyma{\'{n}}ski}}\ and\ \bibinfo {author} {\bibfnamefont {K.}~\bibnamefont
  {{\.{Z}}yczkowski}},\ }\href {\doibase 10.1088/1751-8121/ab4543} {\bibfield
  {journal} {\bibinfo  {journal} {J. Phys. A: Math. Theor.}\ }\textbf {\bibinfo
  {volume} {53}},\ \bibinfo {pages} {015302} (\bibinfo {year}
  {2019})}\BibitemShut {NoStop}%
\bibitem [{\citenamefont {Abbott}\ \emph {et~al.}(2016)\citenamefont {Abbott},
  \citenamefont {Alzieu}, \citenamefont {Hall},\ and\ \citenamefont
  {Branciard}}]{abbott2016tight}%
  \BibitemOpen
  \bibfield  {author} {\bibinfo {author} {\bibfnamefont {A.~A.}\ \bibnamefont
  {Abbott}}, \bibinfo {author} {\bibfnamefont {P.-L.}\ \bibnamefont {Alzieu}},
  \bibinfo {author} {\bibfnamefont {M.~J.~W.}\ \bibnamefont {Hall}}, \ and\
  \bibinfo {author} {\bibfnamefont {C.}~\bibnamefont {Branciard}},\ }\href
  {https://www.mdpi.com/2227-7390/4/1/8} {\bibfield  {journal} {\bibinfo
  {journal} {Mathematics}\ }\textbf {\bibinfo {volume} {4}} (\bibinfo {year}
  {2016})}\BibitemShut {NoStop}%
\bibitem [{\citenamefont {Schwonnek}\ \emph {et~al.}(2017)\citenamefont
  {Schwonnek}, \citenamefont {Dammeier},\ and\ \citenamefont
  {Werner}}]{schwonnek2017state}%
  \BibitemOpen
  \bibfield  {author} {\bibinfo {author} {\bibfnamefont {R.}~\bibnamefont
  {Schwonnek}}, \bibinfo {author} {\bibfnamefont {L.}~\bibnamefont {Dammeier}},
  \ and\ \bibinfo {author} {\bibfnamefont {R.~F.}\ \bibnamefont {Werner}},\
  }\href {https://journals.aps.org/prl/abstract/10.1103/PhysRevLett.119.170404}
  {\bibfield  {journal} {\bibinfo  {journal} {Phys. Rev. Lett.}\ }\textbf
  {\bibinfo {volume} {119}},\ \bibinfo {pages} {170404} (\bibinfo {year}
  {2017})}\BibitemShut {NoStop}%
\bibitem [{\citenamefont {Giorda}\ \emph {et~al.}(2019)\citenamefont {Giorda},
  \citenamefont {Maccone},\ and\ \citenamefont {Riccardi}}]{giorda2019state}%
  \BibitemOpen
  \bibfield  {author} {\bibinfo {author} {\bibfnamefont {P.}~\bibnamefont
  {Giorda}}, \bibinfo {author} {\bibfnamefont {L.}~\bibnamefont {Maccone}}, \
  and\ \bibinfo {author} {\bibfnamefont {A.}~\bibnamefont {Riccardi}},\ }\href
  {\doibase 10.1103/PhysRevA.99.052121} {\bibfield  {journal} {\bibinfo
  {journal} {Phys. Rev. A}\ }\textbf {\bibinfo {volume} {99}},\ \bibinfo
  {pages} {052121} (\bibinfo {year} {2019})}\BibitemShut {NoStop}%
\bibitem [{\citenamefont {Kimura}(2003)}]{kimura2003bloch}%
  \BibitemOpen
  \bibfield  {author} {\bibinfo {author} {\bibfnamefont {G.}~\bibnamefont
  {Kimura}},\ }\href
  {https://www.sciencedirect.com/science/article/abs/pii/S0375960103009411}
  {\bibfield  {journal} {\bibinfo  {journal} {Phys. Lett. A}\ }\textbf
  {\bibinfo {volume} {314}},\ \bibinfo {pages} {339} (\bibinfo {year}
  {2003})}\BibitemShut {NoStop}%
\bibitem [{\citenamefont {Byrd}\ and\ \citenamefont
  {Khaneja}(2003)}]{byrd2003characterization}%
  \BibitemOpen
  \bibfield  {author} {\bibinfo {author} {\bibfnamefont {M.~S.}\ \bibnamefont
  {Byrd}}\ and\ \bibinfo {author} {\bibfnamefont {N.}~\bibnamefont {Khaneja}},\
  }\href {\doibase 10.1103/PhysRevA.68.062322} {\bibfield  {journal} {\bibinfo
  {journal} {Phys. Rev. A}\ }\textbf {\bibinfo {volume} {68}},\ \bibinfo
  {pages} {062322} (\bibinfo {year} {2003})}\BibitemShut {NoStop}%
\bibitem [{\citenamefont {G{\"u}hne}\ and\ \citenamefont
  {T{\'o}th}(2009)}]{guhne2009entanglement}%
  \BibitemOpen
  \bibfield  {author} {\bibinfo {author} {\bibfnamefont {O.}~\bibnamefont
  {G{\"u}hne}}\ and\ \bibinfo {author} {\bibfnamefont {G.}~\bibnamefont
  {T{\'o}th}},\ }\href
  {https://www.sciencedirect.com/science/article/abs/pii/S0370157309000623}
  {\bibfield  {journal} {\bibinfo  {journal} {Phys. Rep.}\ }\textbf {\bibinfo
  {volume} {474}},\ \bibinfo {pages} {1} (\bibinfo {year} {2009})}\BibitemShut
  {NoStop}%
\bibitem [{\citenamefont {Kaniewski}\ \emph {et~al.}(2014)\citenamefont
  {Kaniewski}, \citenamefont {Tomamichel},\ and\ \citenamefont
  {Wehner}}]{kaniewski2014entropic}%
  \BibitemOpen
  \bibfield  {author} {\bibinfo {author} {\bibfnamefont {J.}~\bibnamefont
  {Kaniewski}}, \bibinfo {author} {\bibfnamefont {M.}~\bibnamefont
  {Tomamichel}}, \ and\ \bibinfo {author} {\bibfnamefont {S.}~\bibnamefont
  {Wehner}},\ }\href@noop {} {\bibfield  {journal} {\bibinfo  {journal} {Phys.
  Rev. A}\ }\textbf {\bibinfo {volume} {90}},\ \bibinfo {pages} {012332}
  (\bibinfo {year} {2014})}\BibitemShut {NoStop}%
\bibitem [{\citenamefont {G{\"u}hne}(2004)}]{guhne2004characterizing}%
  \BibitemOpen
  \bibfield  {author} {\bibinfo {author} {\bibfnamefont {O.}~\bibnamefont
  {G{\"u}hne}},\ }\href
  {https://journals.aps.org/prl/abstract/10.1103/PhysRevLett.92.117903}
  {\bibfield  {journal} {\bibinfo  {journal} {Phys. Rev. Lett.}\ }\textbf
  {\bibinfo {volume} {92}},\ \bibinfo {pages} {117903} (\bibinfo {year}
  {2004})}\BibitemShut {NoStop}%
\bibitem [{\citenamefont {Lov{\'a}sz}(1979)}]{lovasz1979shannon}%
  \BibitemOpen
  \bibfield  {author} {\bibinfo {author} {\bibfnamefont {L.}~\bibnamefont
  {Lov{\'a}sz}},\ }\href
  {https://ieeexplore.ieee.org/abstract/document/1055985} {\bibfield  {journal}
  {\bibinfo  {journal} {IEEE Trans. Inf. Theory}\ }\textbf {\bibinfo {volume}
  {25}},\ \bibinfo {pages} {1} (\bibinfo {year} {1979})}\BibitemShut {NoStop}%
\bibitem [{\citenamefont {Karp}(1972)}]{karp1972reducibility}%
  \BibitemOpen
  \bibfield  {author} {\bibinfo {author} {\bibfnamefont {R.~M.}\ \bibnamefont
  {Karp}},\ }in\ \href
  {https://link.springer.com/chapter/10.1007/978-1-4684-2001-2_9} {\emph
  {\bibinfo {booktitle} {Complexity of computer computations}}}\ (\bibinfo
  {publisher} {Springer},\ \bibinfo {year} {1972})\ pp.\ \bibinfo {pages}
  {85--103}\BibitemShut {NoStop}%
\bibitem [{\citenamefont {Knuth}(1994)}]{knuth1994sandwich}%
  \BibitemOpen
  \bibfield  {author} {\bibinfo {author} {\bibfnamefont {D.~E.}\ \bibnamefont
  {Knuth}},\ }\href
  {https://www.combinatorics.org/ojs/index.php/eljc/article/view/v1i1a1}
  {\bibfield  {journal} {\bibinfo  {journal} {Electron. J. Comb.}\ }\textbf
  {\bibinfo {volume} {1}},\ \bibinfo {pages} {A1} (\bibinfo {year}
  {1994})}\BibitemShut {NoStop}%
\bibitem [{\citenamefont {Xu}(2022)}]{zhenpeng2022comm}%
  \BibitemOpen
  \bibfield  {author} {\bibinfo {author} {\bibfnamefont {Z.-P.}\ \bibnamefont
  {Xu}},\ }\href@noop {} {}\bibinfo {howpublished} {Private communication}
  (\bibinfo {year} {2022})\BibitemShut {NoStop}%
\bibitem [{\citenamefont {Feige}(1997)}]{feige1995randomized}%
  \BibitemOpen
  \bibfield  {author} {\bibinfo {author} {\bibfnamefont {U.}~\bibnamefont
  {Feige}},\ }\href {\doibase 10.1007/bf01196133} {\bibfield  {journal}
  {\bibinfo  {journal} {Combinatorica}\ }\textbf {\bibinfo {volume} {17}},\
  \bibinfo {pages} {79} (\bibinfo {year} {1997})}\BibitemShut {NoStop}%
\bibitem [{\citenamefont {Chudnovsky}\ \emph {et~al.}(2006)\citenamefont
  {Chudnovsky}, \citenamefont {Robertson}, \citenamefont {Seymour},\ and\
  \citenamefont {Thomas}}]{chudnovsky2006strong}%
  \BibitemOpen
  \bibfield  {author} {\bibinfo {author} {\bibfnamefont {M.}~\bibnamefont
  {Chudnovsky}}, \bibinfo {author} {\bibfnamefont {N.}~\bibnamefont
  {Robertson}}, \bibinfo {author} {\bibfnamefont {P.}~\bibnamefont {Seymour}},
  \ and\ \bibinfo {author} {\bibfnamefont {R.}~\bibnamefont {Thomas}},\ }\href
  {https://www.jstor.org/stable/20159988?seq=1} {\bibfield  {journal} {\bibinfo
   {journal} {Ann. of Math.}\ ,\ \bibinfo {pages} {51}} (\bibinfo {year}
  {2006})}\BibitemShut {NoStop}%
\bibitem [{\citenamefont {Xu}\ and\ \citenamefont
  {Schwonnek}(2022)}]{counterexample}%
  \BibitemOpen
  \bibfield  {author} {\bibinfo {author} {\bibfnamefont {Z.-P.}\ \bibnamefont
  {Xu}}\ and\ \bibinfo {author} {\bibfnamefont {R.}~\bibnamefont {Schwonnek}},\
  }\href@noop {} {}\bibinfo {howpublished} {In preparation} (\bibinfo {year}
  {2022})\BibitemShut {NoStop}%
\bibitem [{\citenamefont {Moroder}\ and\ \citenamefont
  {Gittsovich}(2012)}]{moroder2012calibration}%
  \BibitemOpen
  \bibfield  {author} {\bibinfo {author} {\bibfnamefont {T.}~\bibnamefont
  {Moroder}}\ and\ \bibinfo {author} {\bibfnamefont {O.}~\bibnamefont
  {Gittsovich}},\ }\href
  {https://journals.aps.org/pra/abstract/10.1103/PhysRevA.85.032301} {\bibfield
   {journal} {\bibinfo  {journal} {Phys. Rev. A}\ }\textbf {\bibinfo {volume}
  {85}},\ \bibinfo {pages} {032301} (\bibinfo {year} {2012})}\BibitemShut
  {NoStop}%
\bibitem [{\citenamefont {Seevinck}\ and\ \citenamefont
  {Uffink}(2007)}]{seevinck2007local}%
  \BibitemOpen
  \bibfield  {author} {\bibinfo {author} {\bibfnamefont {M.}~\bibnamefont
  {Seevinck}}\ and\ \bibinfo {author} {\bibfnamefont {J.}~\bibnamefont
  {Uffink}},\ }\href {\doibase 10.1103/PhysRevA.76.042105} {\bibfield
  {journal} {\bibinfo  {journal} {Phys. Rev. A}\ }\textbf {\bibinfo {volume}
  {76}},\ \bibinfo {pages} {042105} (\bibinfo {year} {2007})}\BibitemShut
  {NoStop}%
\bibitem [{\citenamefont {Morelli}\ \emph {et~al.}(2022)\citenamefont
  {Morelli}, \citenamefont {Yamasaki}, \citenamefont {Huber},\ and\
  \citenamefont {Tavakoli}}]{morelli2022entanglement}%
  \BibitemOpen
  \bibfield  {author} {\bibinfo {author} {\bibfnamefont {S.}~\bibnamefont
  {Morelli}}, \bibinfo {author} {\bibfnamefont {H.}~\bibnamefont {Yamasaki}},
  \bibinfo {author} {\bibfnamefont {M.}~\bibnamefont {Huber}}, \ and\ \bibinfo
  {author} {\bibfnamefont {A.}~\bibnamefont {Tavakoli}},\ }\href {\doibase
  10.1103/PhysRevLett.128.250501} {\bibfield  {journal} {\bibinfo  {journal}
  {Phys. Rev. Lett.}\ }\textbf {\bibinfo {volume} {128}},\ \bibinfo {pages}
  {250501} (\bibinfo {year} {2022})}\BibitemShut {NoStop}%
\bibitem [{\citenamefont {Benatti}\ \emph {et~al.}(2004)\citenamefont
  {Benatti}, \citenamefont {Floreanini},\ and\ \citenamefont
  {Piani}}]{benatti2004non}%
  \BibitemOpen
  \bibfield  {author} {\bibinfo {author} {\bibfnamefont {F.}~\bibnamefont
  {Benatti}}, \bibinfo {author} {\bibfnamefont {R.}~\bibnamefont {Floreanini}},
  \ and\ \bibinfo {author} {\bibfnamefont {M.}~\bibnamefont {Piani}},\ }\href
  {https://www.worldscientific.com/doi/abs/10.1007/s11080-004-6622-6}
  {\bibfield  {journal} {\bibinfo  {journal} {Open Syst. Inf. Dyn.}\ }\textbf
  {\bibinfo {volume} {11}},\ \bibinfo {pages} {325} (\bibinfo {year}
  {2004})}\BibitemShut {NoStop}%
\bibitem [{\citenamefont {G\"uhne}\ \emph {et~al.}(2006)\citenamefont
  {G\"uhne}, \citenamefont {Mechler}, \citenamefont {T\'oth},\ and\
  \citenamefont {Adam}}]{guehne2006entanglement}%
  \BibitemOpen
  \bibfield  {author} {\bibinfo {author} {\bibfnamefont {O.}~\bibnamefont
  {G\"uhne}}, \bibinfo {author} {\bibfnamefont {M.}~\bibnamefont {Mechler}},
  \bibinfo {author} {\bibfnamefont {G.}~\bibnamefont {T\'oth}}, \ and\ \bibinfo
  {author} {\bibfnamefont {P.}~\bibnamefont {Adam}},\ }\href {\doibase
  10.1103/PhysRevA.74.010301} {\bibfield  {journal} {\bibinfo  {journal} {Phys.
  Rev. A}\ }\textbf {\bibinfo {volume} {74}},\ \bibinfo {pages} {010301}
  (\bibinfo {year} {2006})}\BibitemShut {NoStop}%
\bibitem [{\citenamefont {G{\"u}hne}\ and\ \citenamefont
  {Lewenstein}(2004)}]{guhne2004entropic}%
  \BibitemOpen
  \bibfield  {author} {\bibinfo {author} {\bibfnamefont {O.}~\bibnamefont
  {G{\"u}hne}}\ and\ \bibinfo {author} {\bibfnamefont {M.}~\bibnamefont
  {Lewenstein}},\ }\href
  {https://journals.aps.org/pra/abstract/10.1103/PhysRevA.70.022316} {\bibfield
   {journal} {\bibinfo  {journal} {Phys. Rev. A}\ }\textbf {\bibinfo {volume}
  {70}},\ \bibinfo {pages} {022316} (\bibinfo {year} {2004})}\BibitemShut
  {NoStop}%
\bibitem [{\citenamefont {Costa}\ \emph {et~al.}(2018)\citenamefont {Costa},
  \citenamefont {Uola},\ and\ \citenamefont {G\"uhne}}]{costa2018steering}%
  \BibitemOpen
  \bibfield  {author} {\bibinfo {author} {\bibfnamefont {A.~C.~S.}\
  \bibnamefont {Costa}}, \bibinfo {author} {\bibfnamefont {R.}~\bibnamefont
  {Uola}}, \ and\ \bibinfo {author} {\bibfnamefont {O.}~\bibnamefont
  {G\"uhne}},\ }\href {\doibase 10.1103/PhysRevA.98.050104} {\bibfield
  {journal} {\bibinfo  {journal} {Phys. Rev. A}\ }\textbf {\bibinfo {volume}
  {98}},\ \bibinfo {pages} {050104} (\bibinfo {year} {2018})}\BibitemShut
  {NoStop}%
\bibitem [{\citenamefont {Cabello}\ \emph {et~al.}(2014)\citenamefont
  {Cabello}, \citenamefont {Severini},\ and\ \citenamefont
  {Winter}}]{cabello2014graph}%
  \BibitemOpen
  \bibfield  {author} {\bibinfo {author} {\bibfnamefont {A.}~\bibnamefont
  {Cabello}}, \bibinfo {author} {\bibfnamefont {S.}~\bibnamefont {Severini}}, \
  and\ \bibinfo {author} {\bibfnamefont {A.}~\bibnamefont {Winter}},\ }\href
  {\doibase 10.1103/PhysRevLett.112.040401} {\bibfield  {journal} {\bibinfo
  {journal} {Phys. Rev. Lett.}\ }\textbf {\bibinfo {volume} {112}},\ \bibinfo
  {pages} {040401} (\bibinfo {year} {2014})}\BibitemShut {NoStop}%
\bibitem [{\citenamefont {Hastings}\ and\ \citenamefont
  {O'Donnell}(2022)}]{hastings2022optimizing}%
  \BibitemOpen
  \bibfield  {author} {\bibinfo {author} {\bibfnamefont {M.~B.}\ \bibnamefont
  {Hastings}}\ and\ \bibinfo {author} {\bibfnamefont {R.}~\bibnamefont
  {O'Donnell}},\ }in\ \href {\doibase 10.1145/3519935.3519960} {\emph {\bibinfo
  {booktitle} {Proceedings of the 54th Annual ACM SIGACT Symposium on Theory of
  Computing}}},\ \bibinfo {series and number} {STOC 2022}\ (\bibinfo
  {publisher} {Association for Computing Machinery},\ \bibinfo {address} {New
  York, NY, USA},\ \bibinfo {year} {2022})\ p.\ \bibinfo {pages}
  {776–789}\BibitemShut {NoStop}%
\end{thebibliography}%
\end{document}